\documentclass[12pt]{article}
\usepackage{epsfig}
\usepackage{amsmath}
\usepackage{graphicx}
\usepackage{caption}
\usepackage{subfigure}

\begin{document}

\begin{center}
  {\Large The Effects of Diffusion of Information on Epidemic Spread \\ A Multilayer Approach}
\end{center}

\begin{center}
  {{
    \sc Semra G\"und\"u\c{c}}$^*$,\\[2pt]
        Department of Computer Engineering \\
        Faculty of Engineering, Ankara University\\
        06345 G\"olbas{\i} Ankara, Turkey\\
$^*${e-mail{gunduc@ankara.edu.tr}}}
\end{center}


\begin{abstract}
{In this work, the aim is to study the spread of a contagious disease
and information on a multilayer social system.  The main idea is to
find a criterion under which the adoption of the spreading information
block or suppress the epidemic spread. A two-layer network is the base
of the model. The first layer describes the direct contact
interactions while the second layer is the information propagation
layer.  Both layers consist of the same nodes. The society consists of
five different categories of individuals: Susceptibles, infective,
recovered, vaccinated and precautioned. Initially, only one infected
individual starts transmitting the infection. Direct contact
interactions spread the infection to the susceptibles. The information
spreads through the second layer. The SIR model is employed for the
infection spread while the Bass equation models the adoption of
information. The control parameters of the competition between the
spread of information and spread of disease are the topology and the
density of connectivity. The topology of the information layer is a
scale-free network with increasing density of edges. In the contact
layer, regular and scale-free networks with the same average degree
per node used interchangeably.  The observation is that increasing
complexity of the contact network reduces the role of individual
awareness.  If the contact layer consists of networks with limited
range connections, or the edges sparser than the information network
spread of information plays a significant role in controlling the
epidemics.
}
{Social Networks, Multilayer networks, Epidemic,
SIR model, Diffusion of Information, Bass Model.}
\end{abstract}

\section{introduction}

Social interactions are complicated relations of competing
interests. In this sense, the social systems are complex systems. The
modelling of the social phenomena requires a good understanding of the
real interaction patterns and the dynamics among the members of the
society. In the last 20 years the intrest on the compex networs has
intensified. Recent realistic models of real-world complex
systems~\cite{Newman:2003,Castellano:2009} has improved the
understanding of a large variety of complex social interactions. Never
the less, as the knowledge changes new aspects of the social
interactions enter the modelling considerations. As the societies
become more technology oriented, new channels of communication and
interaction rapidly changed the structure and the topology of the
interaction networks. Previously designed single layer real-world
networks left their place to multilayer networks: A developing social
phenomenon finds its reflections on the other layers of social
networks as different types of interactions. The spread of contagious
disease is an excellent example of this situation. In the contact
layer, the interactions of the individuals result in the spread of
infection while in the second layer, the information on the
contagiousness of the disease motivates the individuals to take
preventive measures. Considering the severeness of both human and
economy vise results of an epidemic, the importance of the more
in-depth understanding of the role of inter-woven social networks
during an epidemic spread becomes more apparent.

$\;$

The mathematical models of diffusion of contagious diseases have a
long history, starting early 20th century. The early
models~\cite{Bailley:1975, Anderson:1992} are aggregate models. The
early models have rapidly evolved to agent-based models and lately
also incorporated the structure of the underlying social
networks~\cite{Pastor:2001, Pastor:2015}. The
epidemic studies on networks~\cite{Moreno:2002,Newman:2002,Chakrabarti:2008, Mieghem:2009} are not only usefull and limited to
the spread of contagious diseases within the human societies. A large
variety of complex systems, such as physical, engineering,
technological, and information networks exhibit similar diffusion of
malicious agents~\cite{Kleinberg:2007, Wang:2009, Meisel:2010,
  Guille:2013}.  Complex networks are potent tools to describe
spreading phenomena in both human societies and the other real-life
problems. Never the less the spreading phenomena among the human
societies have more elements than a single complex network. As a
simple example, the traveling individuals change the dynamics of
spreading infections. Similarly, information networks and social
networks affect the dynamic of spreading.  Hence recently the models
of spreading the infections are extended from single complex network
to multilayer networks. Multilayer networks~\cite{Kivela:2013,
  Boccaletti:2014, Shao:2011} are composed of several layers of
complex structures in which the same node may have multiple channels of
interactions.

In this paper, the focus is on the topic of spreading of contagious
disease while the nodes communicate on the severeness of the
epidemic. A two-layer network is employed. The first layer is the
network of contact interactions where the epidemic spreads, while the
interaction on the second layer spreads the information. The
individual gain awareness by the information gathering from the
information network.

The dynamics of epidemic spreading is one of the hottest research
topics in complex network science. The most commonly encountered
contagious diseases are suitably modeled by the
susceptible-infected-recovered ($SIR$),
susceptible-infected-susceptible ($SIS$) and susceptible-infected
($SI$) Epidemiology models. Different spreading mechanisms and
epidemic control strategies are introduced for all three types of
epidemiology models on complex networks~\cite{Pastor:2001,
  Pastor:2002, Madar:2004, Pastor:2015}.  In fighting the infectious
diseases, best prevention strategy is the immunization. The
immunization of the whole population is not a possibly realizable
challange~\cite{Buono:2014}. Hence various strategies of immunization
which may be effective in the prevention of further spreading the
infection are introduced. Random immunization and targetted (selected)
immunization are the methods which aim to block the spreading paths of
the contagion. The efficiency of the immunization is greater if one
can select highly connected nodes.  Such a selection requires prior
knowledge of the whole network. Another immunization strategy is the
acquaintance immunization in which the selection of highly connected
nodes is naturally realized~\cite{Pastor:2002,Cohen:2003,Madar:2004,Chen:2008,Gao:2011}.

An other approach to the efficient immunization is awareness motivated
immunization: the informed individuals decide to take precautions. 
The most effective element of the decision-making process is the
word-of-mouth. The word-of-mouth immediately recalls one-to-one
interaction. In the real-life, the contact networks are only the small
part of the interaction network. In the modern societies, most of the
information comes from the virtual-communication networks. In this
sense, the  word-of-mouth is all trustable
one-to-one correspondences. The human element of the immunization
strategies constitutes spread of information and decision-making
processes. The multiple networks widen the understanding of epidemic
and epidemic control methods by introducing multiple spread
mechanisms. The best example is the disease spreading on the contact
network during the diffusion of information on another. The
information creates the awareness of which is essential to control the
epidemic spread~\cite{Granell:2013, Granell:2014, Wei:2016} hence, the
competition between the awareness and the disease spreading may rise
to an epidemic treshold~\cite{Wang:2014, Wang:2015, Wang:2016}

In the multilayer networks the same nodes, the constituents of the
society, are shared by different layers of the network. Each layer has
different edge topologies and dynamics.  The multilayer networks
capture correct interaction structures between the nodes since an
individual in a society may have different kinds of interactions such
as business relations, social environment, connections through social
media. Hence each is in direct contact with some members of the
society while communicating with some others on a virtual network of
friends.  On online social networks, the information propagates
between the nodes through friendship connections which may be entirely
different from the contact network of the nodes.

In the proposed model, two interacting networks constitute the base of
the artificial society. The first network is the contact network in
which contact interactions result in the spread of contagious
disease. At this layer, SIR model governs the dynamics of the
diffusion of contagion. The second layer, information spread layer,
connects the same nodes with a different connectivity pattern. Not all
of the informed individuals act upon the received knowledge. There is
an adoption process after which the individual reacts. The Bass
model~\cite{Bass:1969} governs the information adoption process. The
Bass model originally was introduced to describe the adoption process
of a new product.  Despite its simplicity of the model is still
successful to explain the diffusion of new ideas, information and it
is commonly used in marketing studies. The main success of the Bass
Model is due to the well represented social behavior of the
individuals. This classification is based on Roger's seminal
work~\cite{Roger:1962, Rogers:2003} on the diffusion of
innovation. Bass Model assumes to types of individuals. The first type
accepts the new idea as soon as it is introduced. The second one is
the majority of the population who like to see the benefit of adaption
of the new information. In this work the Bass model~\cite{Bass:1969}
sets the dynamics of information spread. In the multilayer approach to
disease spread, adoption of the information may lead to the adoption
of a method of prevention of the disease.

\section{The Model}
\label{Model}

The model consists of $N$ nodes which accommodate $N$ interacting
individuals. A two-layer multiplex network, which has common nodes but
different connectivity pattern, connects individuals with each
other. Two sets of parameters, each set indicating the state of the
individual in the corresponding layer, identify the individuals. Hence
the $i^{\rm th}$ node is represented by
$X_i[S_{layer_1},S_{layer_2}]$.  The first layer is the contact layer
where the infection spreads.  The contact layer state parameter,
$S_{Layer_1}$ has five values: Susceptible $S$, infected $I$,
recovered $R$, vaccinated $V$ and the letter $P$, indicate the
individual who has taken precautinory measures. The individuals who
take precautionary measures remain susceptible, with a reduced
probability of interactions with their neighbors. The information
spreads on the second, virtual, layer. The awareness parameter,
$S_{Layer_2}$ takes only two values aware (informed) $A\!W$ and
non-aware (uninformed) $N\!\!A$.

\subsection{Interactions}
\label{Interactions}

Initially all nodes are initialized as susceptible, $S_{Layer_1}=S$,
and non-aware, $S_{Layer_2}=N\!\!A$. The infection spread from only
one randomly chosen node $S_{Layer_1}=I$.  An infected individual
automatically becomes aware, $S_{layer_2}=A\!W$. Both contamination
and information spread start from this single node. The infection
spreads in the first layer by the contact interactions. SIR model
dynamics,

\begin{eqnarray}
    \label{SIRModel}
\frac{dS(t)}{dt} &=& - \beta \, I(t)\, S(t) \nonumber \\
\frac{dI(t)}{dt} &=&  \beta \, I(t)\, S(t) - \gamma\, I(t)  \\
\frac{dR(t)}{dt} &=& \gamma \, I(t)\nonumber \\
\end{eqnarray}

where $S(t)$, $I(t)$, and $R(t)$ are the number of susceptible,
infected and removed individuals at time $t$.  The SIR model
has two free parameters, $\beta$ and $\gamma$.  The
parameter, $\beta$, represents an average rate of encounters between
the infected and susceptible individuals. The second parameter,
$\gamma$ is the rate of recovery per unit time. The recovered infected individuals gain immunity.

For the agent-based simulation model, SIR model dynamics is
implemented as, probabilistic interactions among the members of the
society. At each time step, randomly selected nodes interact with the
neighbors at the contact layer and spread information on the virtual
network. The rules are: If a susceptible or a precautioned individual
interacts with an infected neighbor, become infected ($S \;\;{\rm and}
\;\;P \rightarrow I$) with probability $\beta$. Recovered, $R$ and
vaccinated, $V$ individuals are not affected from an infected member
of the society. The precautioned individuals, $P$ avoid interaction
with individuals in any state with probability $Prb$.

The information layer serves for two purposes: Spread and adoption of
the information on the disease. For both of these processes Bass model
is suitable since the model parametrize the human behavior for the
adoption processes. The original form of the Bass model assumes two
different type of individuals. The first group is the innovators who
adopt a new idea immediately after being informed. The second group is
the imitators who want to see the results of the accepting the new
idea by observing the results on already adopted individuals.  A new
idea starts to diffuse through innovators. After a certain number of
initial adopters, imitators are the main driving force in the spread
of information. In the classical form the Bass equation,

\begin{equation}
  \label{BassModel}
\frac{dA\!W(t)}{dt} = (p + \frac{q}{N} A\!W(t)  ) (N - A\!W(t))
\end{equation}

where, $p$ and $q$ are innovation and imitation parameters, $N$ and
$A\!W$ are the total number and the number of aware individuals. Here,
the innovation parameter best understood as the probability of
adoption of new information immediately after being informed. The
imitations parameter is related to the probability of adoption after
observing the experiences of the neighbors (Word-of-mouth). Informed
individuals transmit the information to their neighbors through their
connections on the second layer. When an individual receives the
information evaluates the information.  According to the dynamics
defined by the Bass equation, the information is adopted or not. In
the agent-based approach, the information adoption takes the following
form:

\begin{eqnarray}
{\rm if}\;\; p > r\;\;\; && X_i[S,N\!\!A]=X_i[S,A\!W]\nonumber \\
{\rm else \;if} \;\; \frac{q}{N\!\!N}\times N\!\!N_{A\!W}>  r \;\;\;&& X_i[S,N\!\!A] = X_i[S,A\!W]\nonumber
\end{eqnarray}

here, ${N\!\!N}$ and $N\!\!N_{A\!W}$ are the number of nearest neigbors and number of aware neigbors respectively. 
If the information is adopted the awareness state is set to aware,
$S_{layer_2}=A\!W$. Once an individual is informed remains informed,
but only once sends the message to all neighbors after adopting the
information.

The individuals take precursory measures according to their
attitudes. Two precursory measures are vaccination and reducing the
probability of interactions. If the susceptible individual is
vaccinated ($V$), gain immunity. If an individual takes a precursory
measure of reducing the number of interactions does not gain immunity.
They remain susceptible, but their  interaction
probability is reduced. The interaction at the information layer leads to the
adoption of information and decision of a precursory action.

If an individual is susceptible and informed, $ X_i[S,A\!W]$, may take
precaution or may prefer vaccination. 

\begin{eqnarray}
{\rm if}\;\; X_i[S,A\!W]=\left\{\begin{array}{ll}X_i[P,A\!W] & {\rm if} \;\; Prb > r \\  X_i[V,A\!W] & {\rm if} \;\; (1 - Prb) > r \end{array}\right. \nonumber \\
\end{eqnarray}

The unaffected and aware individuals may be in two states: Vaccinated,
$V$, or precaution is taken, $P$ in which case the probability of
interactions of the individual changes.  At each time step a randomly
choosen individual interact with a randomly choosen neighbor. The
individual and its neighbor can be in any of the five states. Unles
the interaction is between a susceptible and contaminated, interacting
individulas do not change state. There are two types of susceptibles:
$S$ and $P$ state individuals.  For $S$ state each interaction with a
infective individual spread contamination. For the individuals who are
in the $P$ state, the individual does not interact at every time step
even if they are choosen.  Their interactions are limited with a
probability $p_{interaction}$ which represent the prevention effort of
the $P$-state individuals. Interaction probability is kept constant as
$p_{interaction}=0.25$, only one fourth of the encountrs ends with a
physical contact. If a $P$ state individual interact with an $I$ state
individual change state.

\subsection{The multiplex network}
\label{Network}

Two interconnected networks, one for contact interactions and the
second one for the spread of information carry the social
interactions.  Both systems share the same nodes with different
intra-layer connectivity structures. The proliferation of contagious
disease progresses on the contact network. The contact network layer
has two alternative network structures: Regular two-dimensional
lattices with periodic boundary conditions and scale-free
networks. The underlying network structure is scale-free for the
information layer. This choice is due to the similarities between the
scale-free and the real-world social network structures.

Both regular and scale-free networks are used as the contact layer. In
the regular network case, periodic boundary condition with simple
square ($k=4$) and triangular ($k=6$) lattices are used to test the
effects of connectivity. Two different scale-free networks with the
same average connectivity ($<k>=4$ and $6$) per node are tested on the
contact interaction layer.  Barab\'asi-Albert network algorithm is
used to generate the scale-free networks. In this algorithm, the
number of seed nodes, $m$, guarantees the average number of undirected
edges, $<k>=2\times m$. Changing the number of seed nodes controls the
density of the number of connections, the degree of the node. The degree
distribution of the nodes affects the spread of the information and
the contagious disease. On the information layer, only scale-free networks are used.  The networks with a wide range of average
degree distributions are obtained by using the Barabasi-Albert algorithm
for the information layer. The effects of information spread on the
spread of contagious disease are tested by using lattices in the range
of $<k>=4$ to $20$ The relation between the connectivity structure of
two layers and the speed of the disease spread is the subject of the
next section.

\section{Results and Discussions}

An artificial society of $N=10000$ inhabitants, each occupying a node
on a multiplex network, are the constituents of the simulation system.
The connectivity of the nodes is two-fold. The first layer of the
multiplex network is the contact network where individuals interact
with each other through direct contact interactions. Hence, the
contact layer provides a media for the transmission of contagion
disease. The second layer is the information layer, through which the
information spread via virtual contacts. The conditions and the speed
of the spread of news and infection are a function of both topology
and the average degree of the nodes. The contact layer consists of
both regular lattice and scale-free networks while for the information
spread layer, only scale-free networks with varying average degree per
node are used. The presented results are the averages of $100$
simulations each starting from a statistically independent initial
configuration. The creation of an initial configuration consits of
creation of multiplex network, initilizing both contact and
information layer state parameters of each node. Iterations continued
until the stationary configurations are reached. The required time
duriation varies according to the topology and the density of the
links of the contact layer. For regular lattices, approximately $250$
time steps are observed to be sufficient. Barab\'asi-Albert network
provides a faster transmiting media.  The system reach the stable
configurations after only $50$ time steps.  During the simulation all
parameters, apart from the lattice parameters are kept fixed to
compare the effects of the lattice topology. The contact layer
parameters which controls the spread of contagious disease, the
infection transmission, $\beta$ and recovery, $\gamma$ parameters of
the SIR model are kept constant for all networks. The transmission and
recovery parameters are $\beta = 1$ and $\gamma=0.2$ respectively.
The information adoption is controled by the Bass equation parameters,
$p$ and $q$.  Individuals who adopt an information immediately after
being informed are rare. The majority adopt after obrerving the
results of fist hand experiances.  The values of innovation and
immitation parameters are assumed to be similar to those of the
average values obtained from the marketing studies. From marketing the
average ranges are $0.001 < p < 0.1$ and $0.1 < q < 0.5$ for
innovation and imitation parameters respectively.  In this work, the
fixed values of $p=0.05$ and $q = 0.35$ are employed for all lattices.

In the societies, the direct contact networks usually have relatively
small average degree per node.  Hence in the contact layer, the
average degree per node is limited to $<k>=4$ and $<k>=6$. The regular
networks are 2-dimensional simple square ($k=4$) and triangular
($k=6$) lattices with periodic boundary conditions, while the
scale-free networks are generated by using Barab\'asi-Albert algorithm
with initial sites of $2\;{\rm and}\;3$ which corresponds to the
average degree, $<k>=4 \;{\rm and}\; 6$. The information spread layer
is expected to have denser connections between the nodes. Hence,
undirected scale-free networks with increasing density of the edges
are generated by using Barab\'asi-Albert algorithm.  The average
degree, $<k>$, per node is the control parameter of the spreads on the
different information networks.
 
\begin{figure}
\subfigure[SIR model on regular networks network \label{Fig:Regular46}]{%
  \includegraphics[height=6cm,width=.49\linewidth]{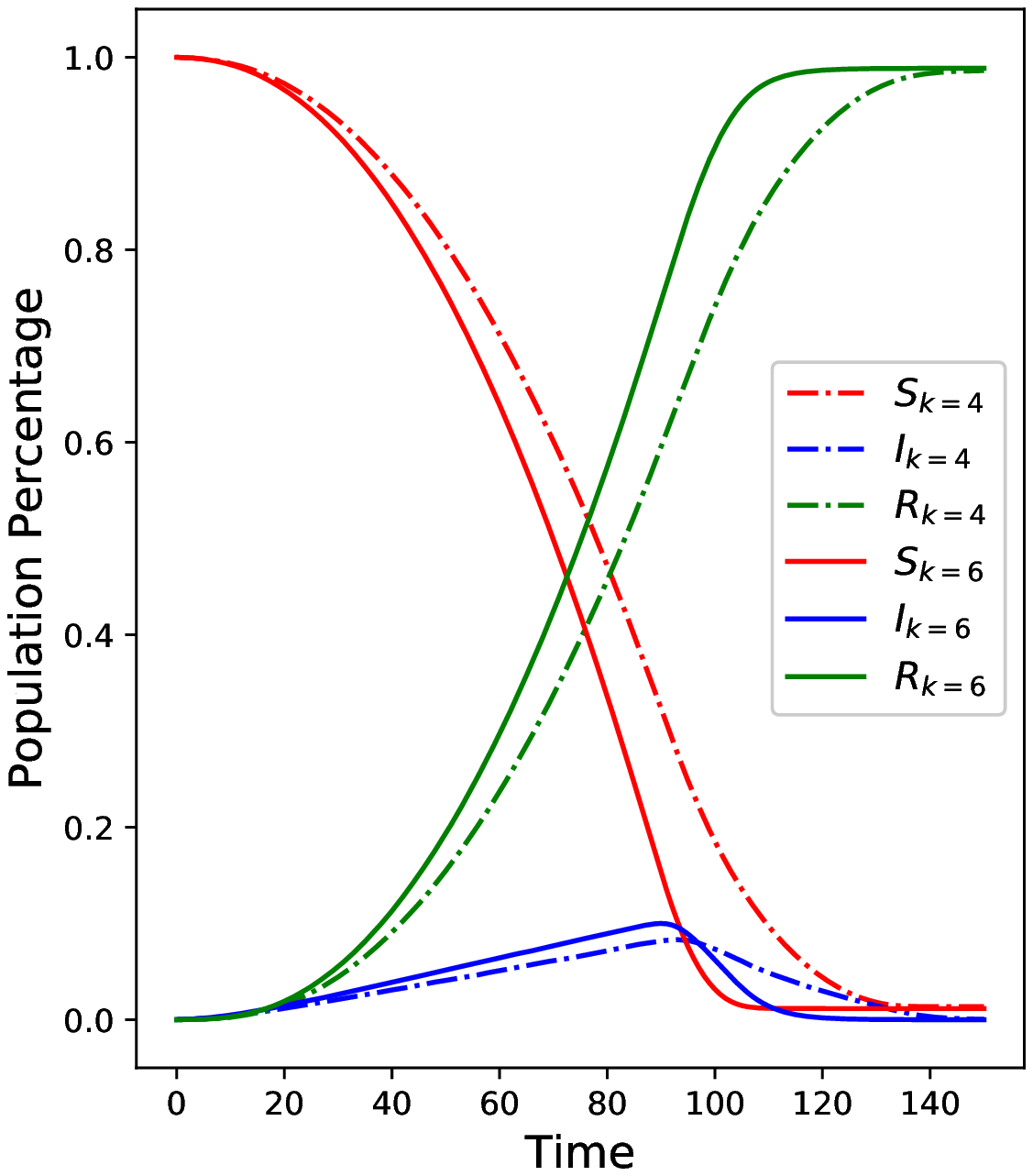}%
}\hfill
\subfigure[SIR model on scale-free network\label{Fig:ScaleFree23}]{%
  \includegraphics[height=6cm,width=.49\linewidth]{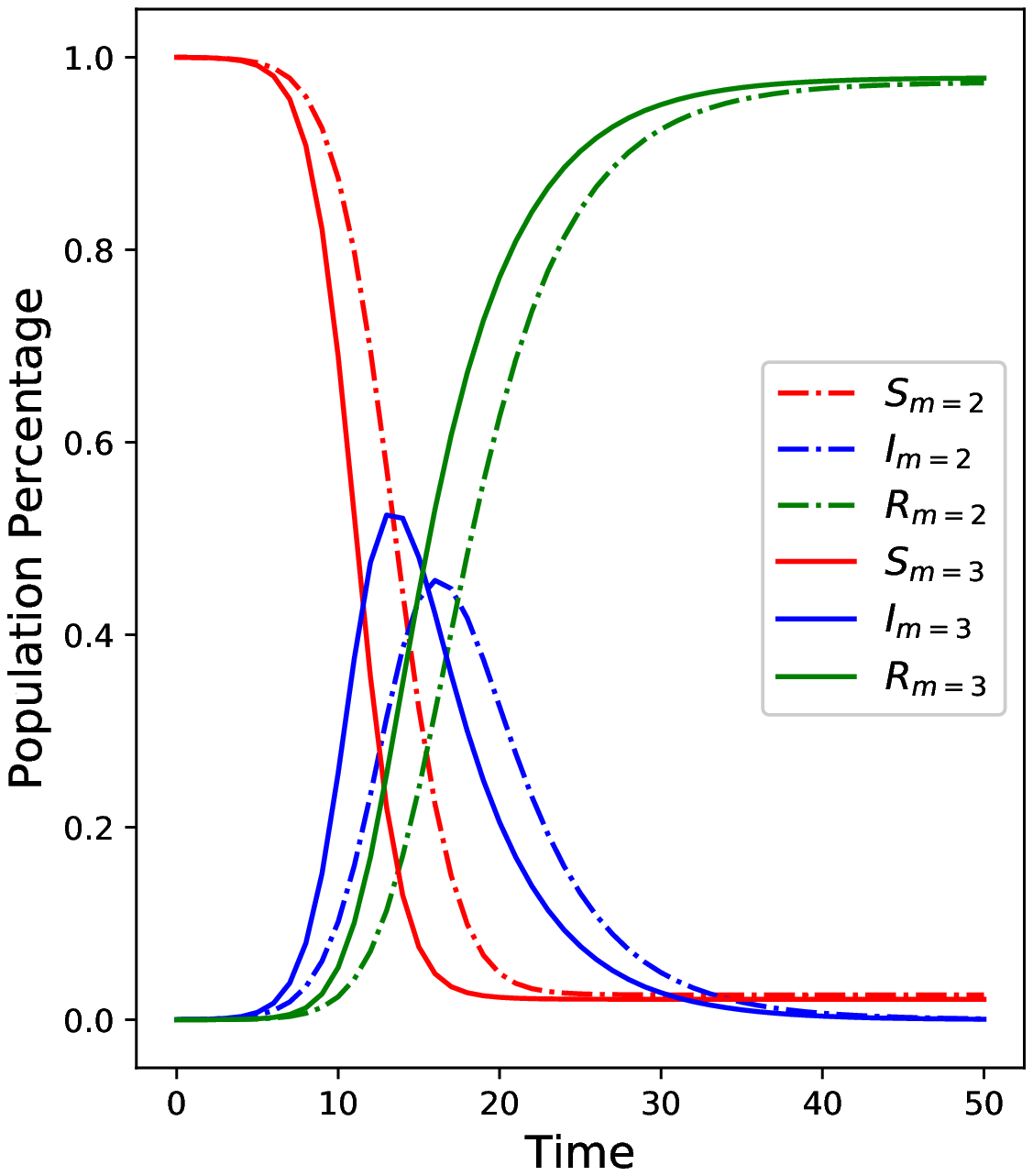}%
}
\caption{Spread of contegious disease in a society with regular and scale-free contact layer network topologies. No information propagation is considered. }
\label{Fig:SimpleSIR}
\end{figure}

Figure~\ref{Fig:SimpleSIR} shows the spread of contagious disease on
regular and scale-free networks without the contribution of
information layer.  Two different topologies, with the same average
degree per node, are square and triangular lattices and $m=2$ and
$m=3$ Barab\'asi-Albert networks respectively.  The comparison between
the figures \ref{Fig:Regular46} and \ref{Fig:ScaleFree23} shows that
the spread of infection on the contact layer, almost three times
faster on the scale-free network than the corresponding regular
network for the same transmission and recovery parameters. Since the
transmission parameter is high, disease spread among the population,
but the peak values of the number of infected individuals are differ
depending on the topology. In the scale-free network case, almost half
of the population is contaminated at the peak of the infection
spread. In the regular lattice case, the peak value of the number of
infected remains around $10 \%$ of the total population. The
increasing number of average connections per node pronaunce the
difference. Its significance becomes more apparent when the
availability of the work-force and continuity of the social system is
considered. Hence reducing the peak value of the number of infected
through mass media and social media plays a crucial role in the
continuity of the social systems.

\begin{figure}
\subfigure[$k=4,\; m=2$ \label{Fig:R4SF02}]{%
  \includegraphics[height=6cm,width=.49\linewidth]{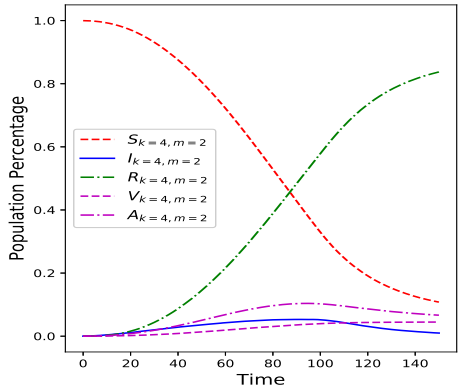}%
}\hfill
\subfigure[$k=4,\; m=4$\label{Fig:R4SF04}]{%
  \includegraphics[height=6cm,width=.49\linewidth]{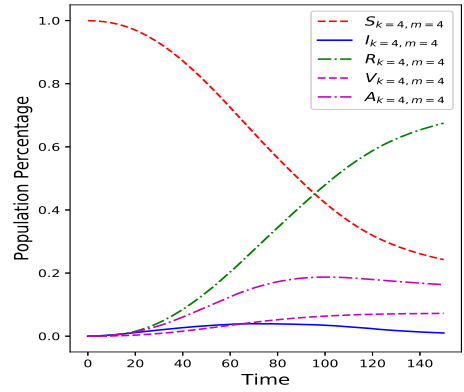}%
}\\

\subfigure[$k=4,\; m=6$ \label{Fig:R4SF06}]{%
  \includegraphics[height=6cm,width=.49\linewidth]{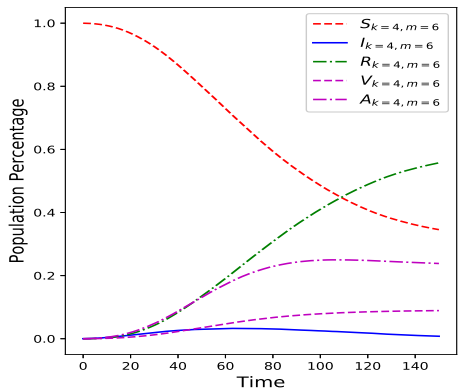}%
}\hfill
\subfigure[$k=4,\; m=10$\label{Fig:R4SF10}]{%
  \includegraphics[height=6cm,width=.49\linewidth]{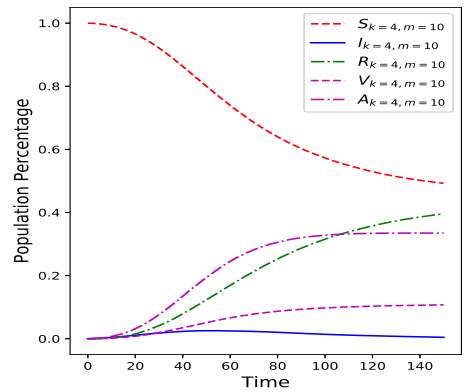}%
}
\caption{Contegious disease spread on simple square lattice, $k=4$,
  while the information spreads on scale-free network topology with
  increasing connectivity.}
\label{fig:R4SF}
\end{figure}

Constructing multiplex networks to study the effects of the
information spread requires comparing multiple information networks
which have the same contact layer network.  As the first set of
examples, a simple square lattice and a set of scale-free networks
with the progressively increasing number of edges are taken as the
topologies of contact and information networks respectively. Figure
\ref{fig:R4SF} shows the effects of increasing number of communication
links. Usually, in the social systems, the contact networks are local
interactions. Hence, diffusion takes more time than real-world
networks. In this first model, increasing number of second layer links
speeds up the spread of information. If the individuals absorb and use
the information in the correct way by taking precautions or getting
vaccinated, the disappearance or at least control of the contamination
is possible. Figures \ref{Fig:R4SF02}, \ref{Fig:R4SF04},
\ref{Fig:R4SF06} and \ref{Fig:R4SF10} show the effects of the
increasing number of links on the spread of contamination. As
explained on the section \ref{Interactions}, the contamination spread
according to the dynamics of the SIR model with an infection
transmission rate, $\beta=1$. If a susceptible contacts with an
infective gets contaminated. Each informs all neighbors when
infected. Adoption of the information is a process governed by the
dynamics provided by the Bass equation. A small percentage of the
individuals (innovators) immediately adopt the information and take a
precaution. Others, collect information from the neighboring nodes
before making a decision. Both innovators and imitators have two
choices as far as the precautions are concerned: getting vaccinated
and avoiding contacts with the neighbors. In this work, the first
assumption is that only $20\%$ of the informed individuals choose
vaccination. The rest prefers to keep away from any contact
interaction. A second assumption is that on the $75\%$ of the
occasions susceptibles can save themselves from contamination by
avoiding direct contact.

\begin{figure}
\subfigure[Susceptible \label{Fig:R4SFS}]{%
  \includegraphics[height=6cm,width=.49\linewidth]{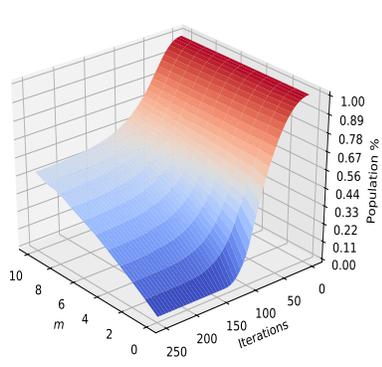}%
}\hfill
\subfigure[Infected\label{Fig:R4SFI}]{%
  \includegraphics[height=6cm,width=.49\linewidth]{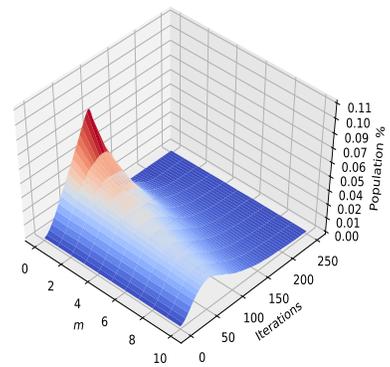}%
}\\

\subfigure[Recovered \label{Fig:R4SFR}]{%
  \includegraphics[height=6cm,width=.49\linewidth]{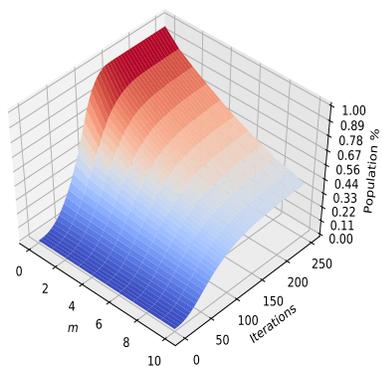}%
}\hfill
\subfigure[Precautioned and vaccinated\label{Fig:R4SFV}]{%
  \includegraphics[height=6cm,width=.49\linewidth]{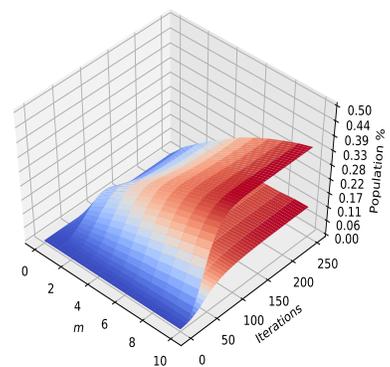}%
}
\caption{The effect of increasing connectivity of the information network on the infection spread. The number of initial sites of the information layer changes from $0$ (network consists of only the contact layer) to $10$. }
\label{Fig:R4SFXX}
\end{figure}

Figure~\ref{Fig:R4SFXX} show the changes in the number of
susceptibles, infected, recovered, vaccinated and precautioned for a
constant speed of contamination spread, the peak of the number of
infected individuals decreases with increasing density of the
information links. Figure \ref{Fig:R4SFS} shows the changes in the
number of susceptibles with as a function of time and number of
initial sites, $m$ (Average degree, $<k>=2*m$ ). For small $m$, all
individuals get infected. As the number of initial sites approaches to
$m=10$, over $40\%$ of all susceptibles remain unaffected from the
contamination which reduces the number of recovered (Figure
\ref{Fig:R4SFR}). Similarly, the peak of the number of infected
individuals decreases rapidly with the increasing number of
information links (Figure \ref{Fig:R4SFI}). The number of vaccinated
remains rather small compared with the number of precaution. Figure
\ref{Fig:R4SFV}, shows the changes in the number of vaccinated (Below)
and the number of precautioned (Above) respect to the changes in the
number of connections in the information layer.

\begin{figure}
\subfigure[ $m_{contact}=2,\; m_{information}=2$ \label{Fig:SF2SF02}]{%
  \includegraphics[height=6cm,width=.49\linewidth]{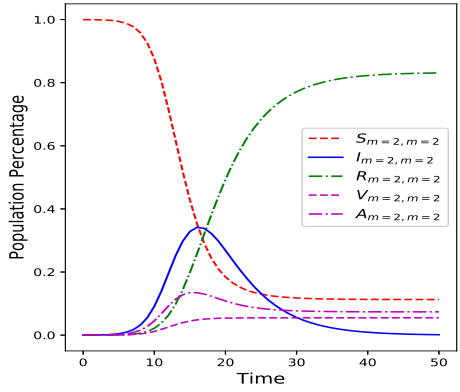}%
}\hfill
\subfigure[$m_{contact}=2,\; m_{information}=4$\label{Fig:SF2SF04}]{%
  \includegraphics[height=6cm,width=.49\linewidth]{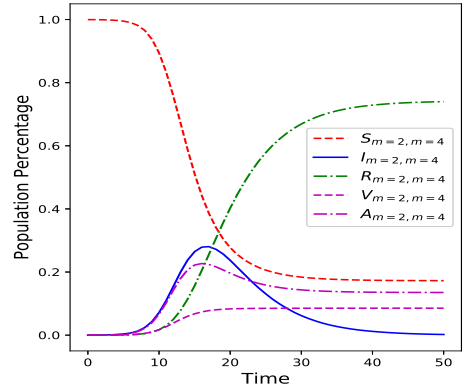}%
}\\

\subfigure[$m_{contact}=2,\; m_{information}=6$ \label{Fig:SF2SF06}]{%
  \includegraphics[height=6cm,width=.49\linewidth]{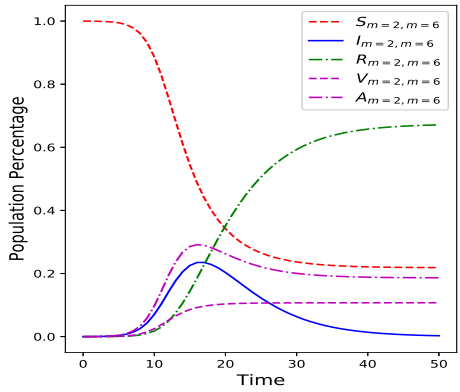}%
}\hfill
\subfigure[$m_{contact}=2,\; m_{information}=10$\label{Fig:SF2SF10}]{%
  \includegraphics[height=6cm,width=.49\linewidth]{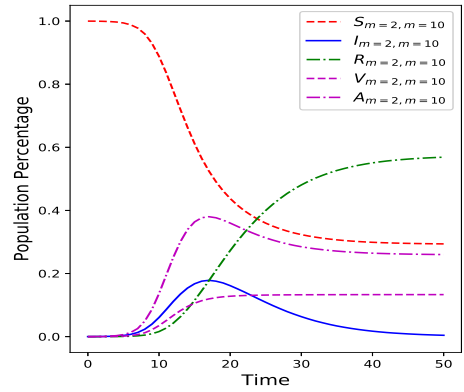}%
}

\caption{Spread of contegious disease in a society with scale-free multilayer network topology. Contact layer has single initial sites configularion, $m=2$ while the number of initial sites of the information layer changes, $m=2,4,6$ and $10$. }
\label{Fig:SF2SF}
\end{figure}

This effect manifests itself more profoundly in scale-free networks in
the contact layer case. When the contact layer is in scale-free
topology, the speed of transmission of infection is high comparing
with the regular networks. Therefore, the peak of the number of
infected is higher in the scale-free contact network concerning
regular networks.  Figures \ref{Fig:Regular46} and
\ref{Fig:ScaleFree23} show the differences in speed and the scale of
the contamination between lattices and scale-free networks with equal
average degree per node. The topology of the information layer plays a
very important role in reducing both the total number of infected
individuals and the peak in the number of infected individuals. Figure
\ref{Fig:SF2SF}, show the effect of the increasing density of
information links while contact network is also scale-free with an
average degree of $4$ per node. The real-world networks, due to
complex connectivity structure, speed up the spreading
phenomena. Figures \ref{Fig:SF2SF02}, \ref{Fig:SF2SF04},
\ref{Fig:SF2SF06}, \ref{Fig:SF2SF10} show the effect of the density of
information layer connections for fixed average degree in the contact
layer. Increasing the number of average degree decrease the peak of
the number of infected. Two effects contribute to the decrease in the
infectives, vaccination and awareness. Informed individuals either get
vaccinated and gain immunity or avoid direct contacts with
neighbors. As the number of communication links increase, the number
of aware individuals increase which result in reducing the number of
infected. Comparison of the figures \ref{Fig:SF2SF02},
\ref{Fig:SF2SF04}, \ref{Fig:SF2SF06}, \ref{Fig:SF2SF10} indicate that
the main contribution in the prevention of the epidemic spread comes
from the group of individuals who try to avoid direct contacts with
the neighbours. This group of individuals increase as the number of
infected individuals increase. Their peak is just before the peak of
the number of infected individuals which prevent further
contamination. As the number of links of the information layer
increase, the peak of informed individuals increases with a further
suppression on the spread of infection. The contribution on the
prevention of the vaccinated does not grow with the same rate.

\begin{figure}
\subfigure[$m_{contact}=3,\; m_{information}=2$ \label{Fig:SF3SF02}]{%
  \includegraphics[height=6cm,width=.49\linewidth]{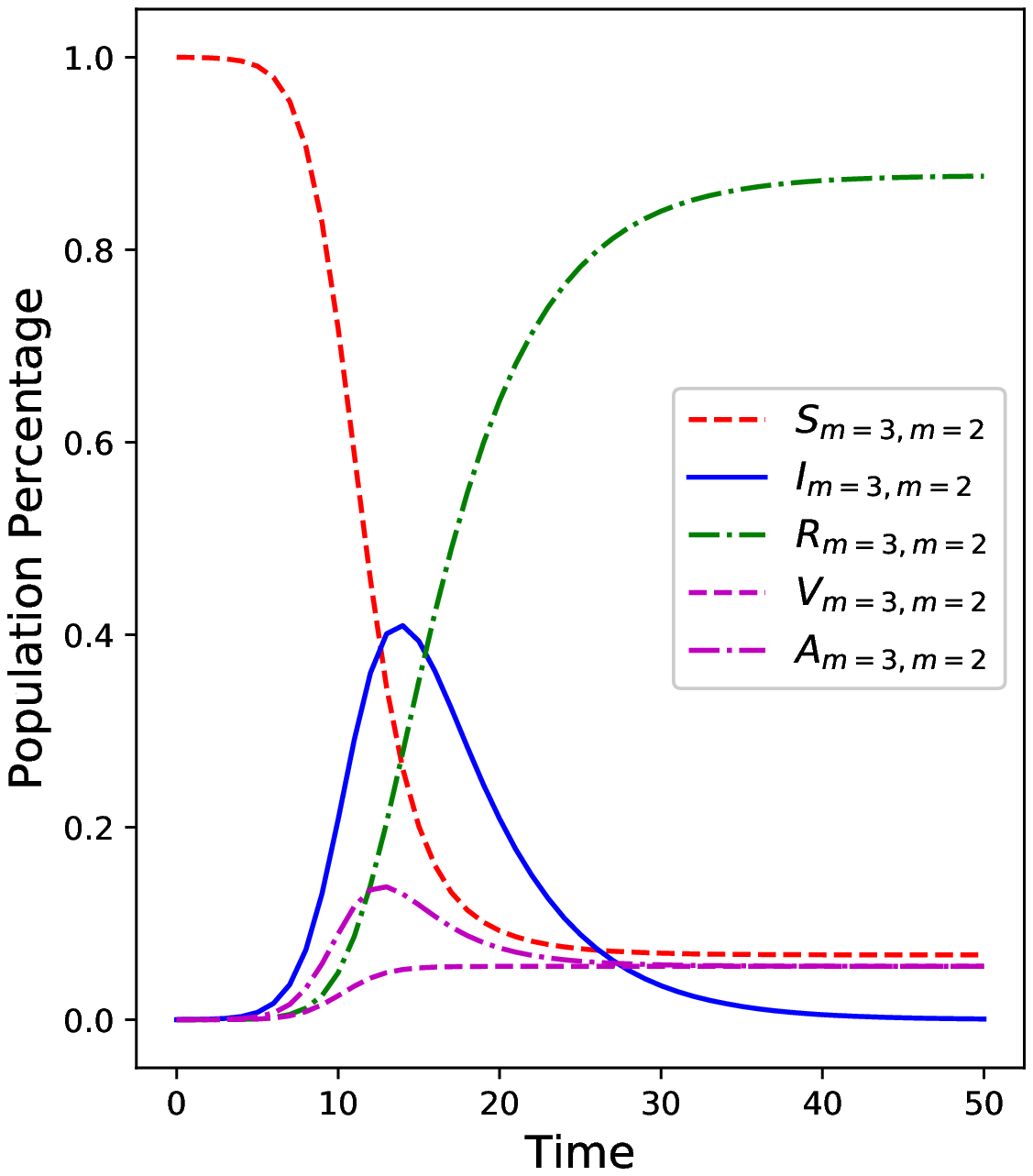}%
}\hfill
\subfigure[$m_{contact}=3,\; m_{information}=4$\label{Fig:SF3SF04}]{%
  \includegraphics[height=6cm,width=.49\linewidth]{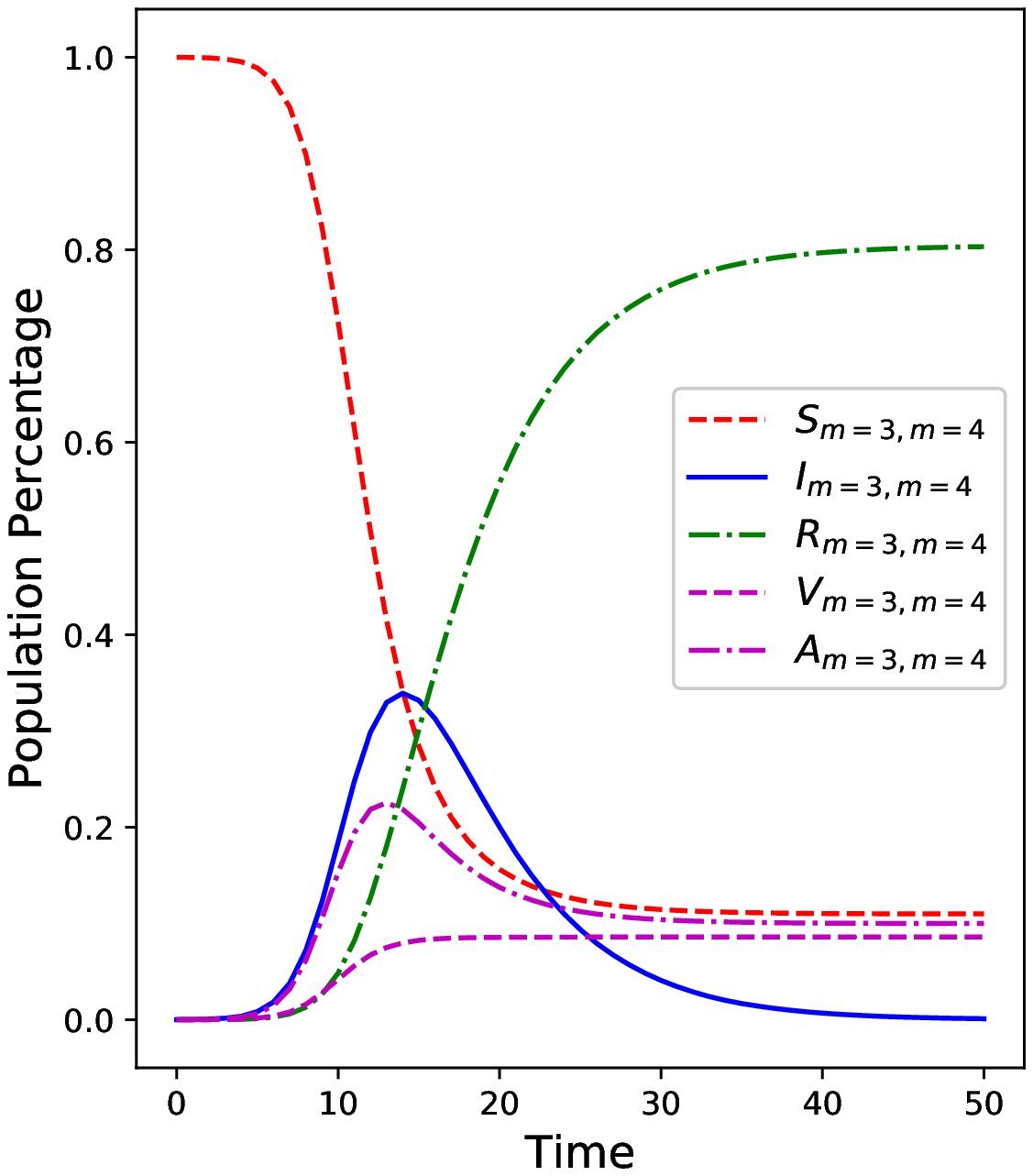}%
}\\

\subfigure[$m_{contact}=3,\; m_{information}=6$ \label{Fig:SF3SF06}]{%
  \includegraphics[height=6cm,width=.49\linewidth]{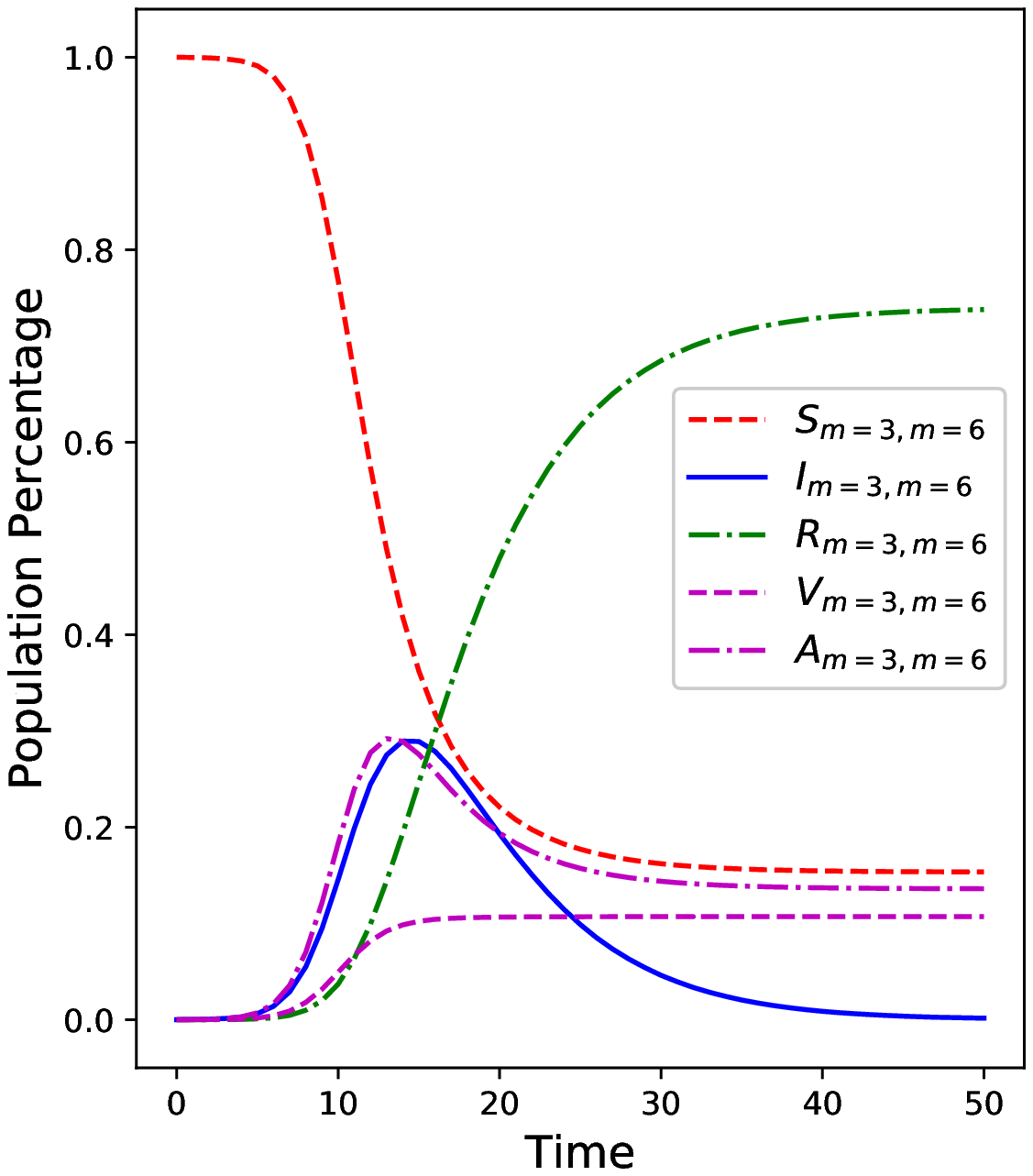}%
}\hfill
\subfigure[$m_{contact}=3,\; m_{information}=10$\label{Fig:SF3SF10}]{%
  \includegraphics[height=6cm,width=.49\linewidth]{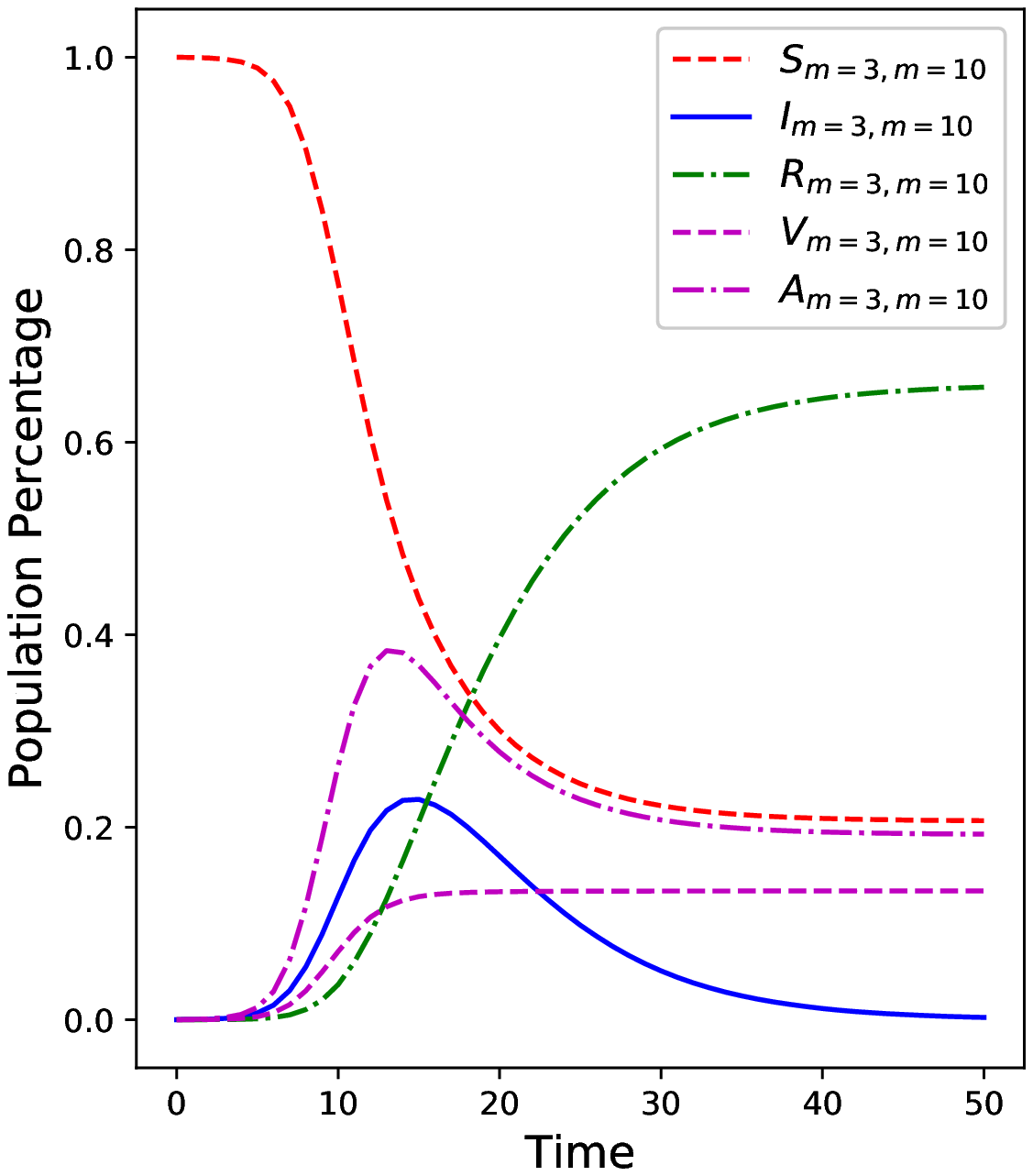}%
}

\caption{Spread of contegious disease in multi-layer network with scale-free network topology. The same as figure~\ref{Fig:SF2SF} only the contact layer has denser connectivity structure, $m=3$.}
\label{Fig:SF3SF}
\end{figure}

When the contact layer becomes denser, the propagation of the
infection is very fast. Hence spread of information to prevent further
spreas of the ilmess is less effective. Figure \ref{Fig:SF3SF} shows
the effect of information spread while the contact layer has
scale-free topology with average degree per node is $6$.

\begin{figure}
\subfigure[Contact layer:Square lattice ($k=4$) \label{Fig:RvsSIR}]{%
  \includegraphics[height=6cm,width=.49\linewidth]{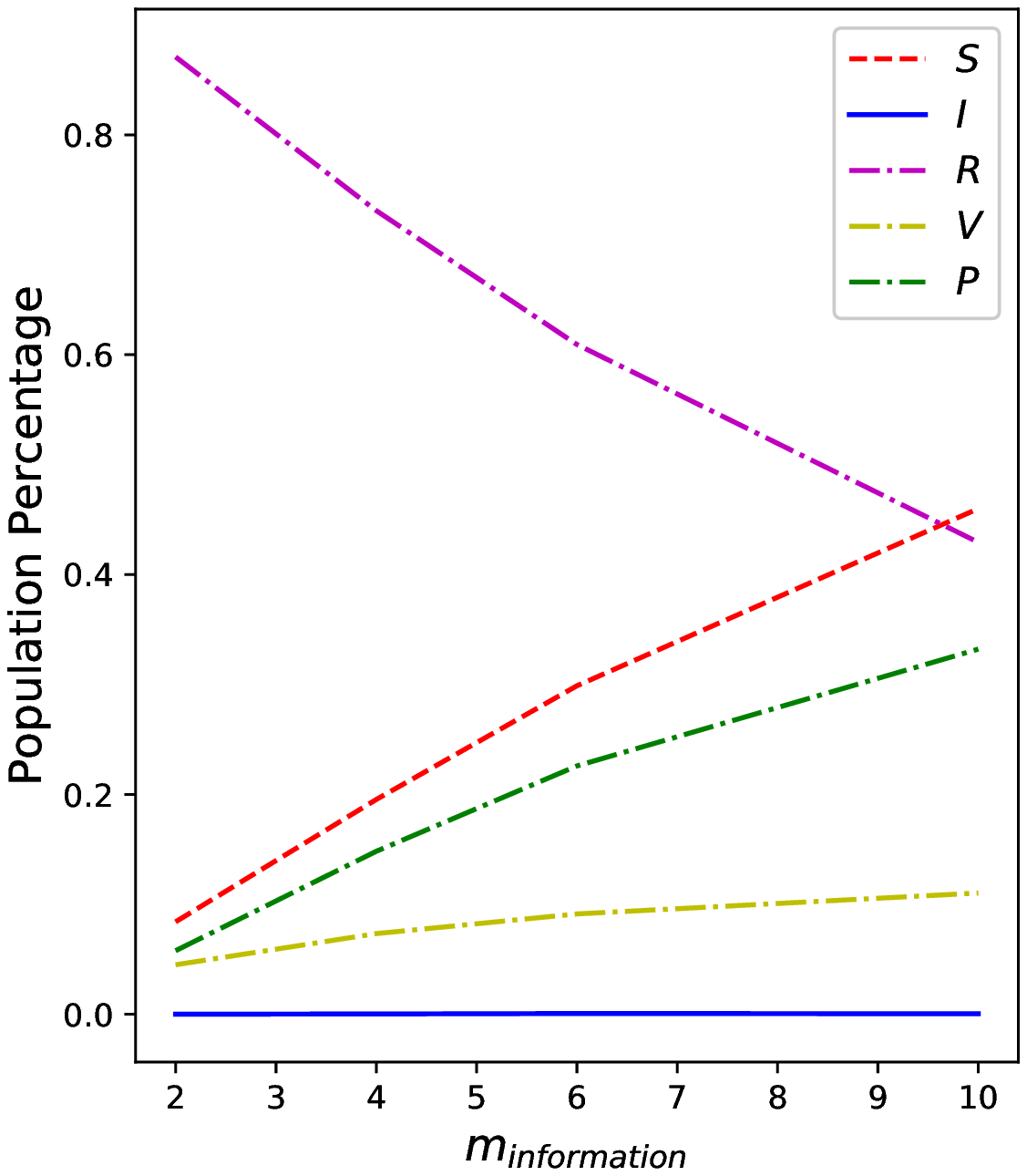}%
}\hfill
\subfigure[Contact layer:Scale-free network ($m=2$)\label{Fig:SFvsSIR}]{%
  \includegraphics[height=6cm,width=.49\linewidth]{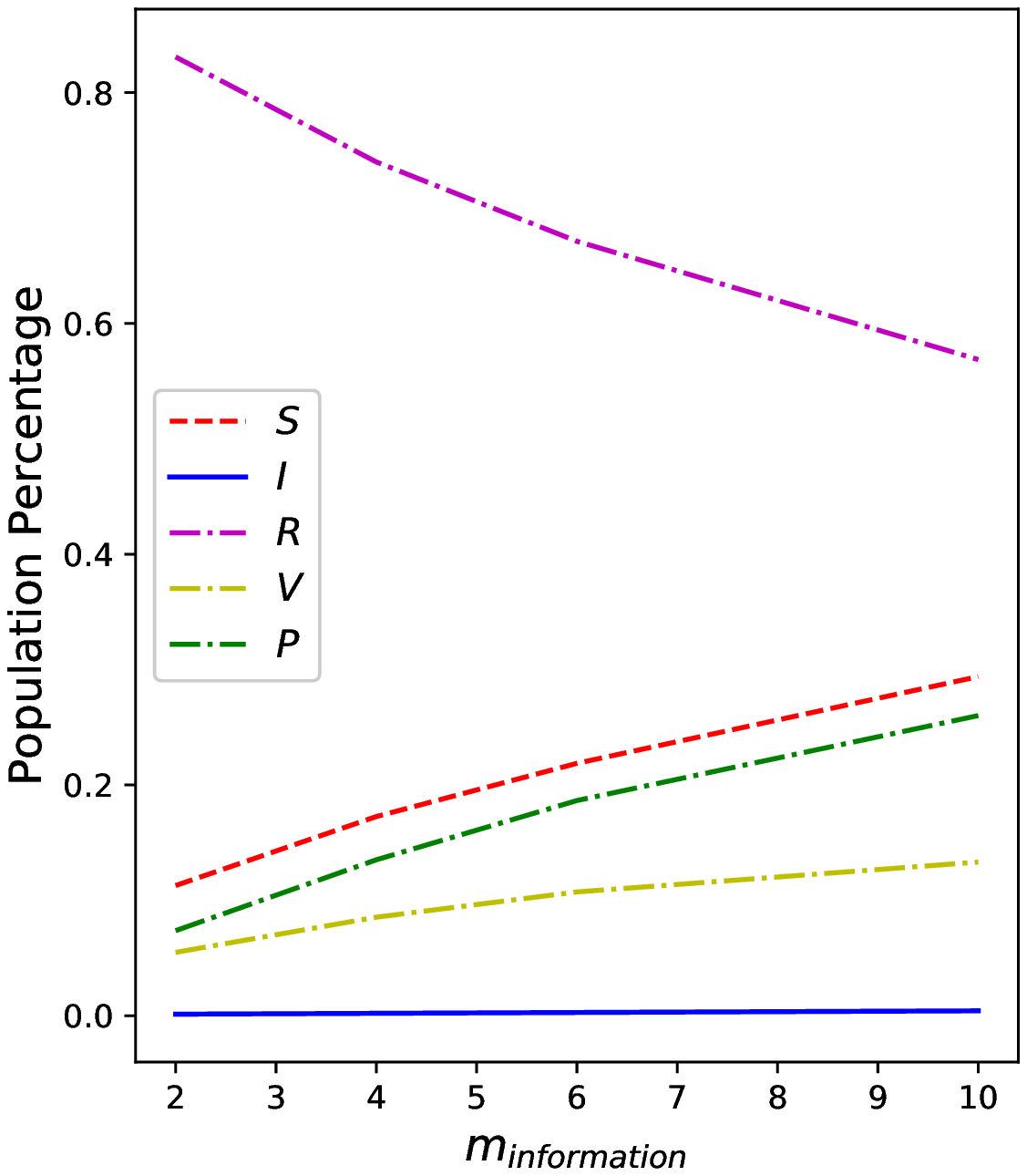}%
}
\caption{Aftermath of the epidemic. The  susceptible, infected, recovered, vaccinated and precautioned population versus the number of initial sites of the information layer. Constant contact layer parameter is fixed to $k=4$ for regular network (a) and $m=2$ for scale-free network (b).   }
\label{Fig:SFSF}
\end{figure}

Figure \ref{Fig:SFSF} summarize the results of the model. The effect
of the density of links on the information layer is observed on two
different contact network topologies, regular and scale-free networks
with equal average degrees per node.  The figures \ref{Fig:RvsSIR} and
\ref{Fig:SFvsSIR} show the percentages of recovered (dash-dot),
susceptible (dashed), precautioned (solid) and vaccinated (dashed)
individuals after the ending the spread of infection. In fact, the
bottom line shows no infected individuals . Figure \ref{Fig:RvsSIR}
indicate that as the average number of links of the information layer
increase the number of infected individuals ($R$) decrease to almost
$40\%$ of the population which indicate that the total number of
healthy (vaccinated or uninfected) reach up to $60 \%$. The situation
changes slightly in the case of scale-free contact layer with
$<k>=4$. Figure \ref{Fig:SFvsSIR} show that with increasing
information spread the percentage of the recovered individuals goes
down to only $60 \%$ of the population. Total percentage of the non
effected individuals are almost $40\%$. The difference in the spread
speed of the infection and information can explain this drastic
difference between two layer topologies on the scale-free networks.


\section{Discussions and Conclusions}

Recent analytical and simulation models indicate that the epidemic
spreading on physical contact networks ignite the spread of
awareness. The awareness of the individuals, in turn, suppress the
disease spreading. In this work, the discussion is the relation
between the epidemic spread and the effect of individual awareness. In
the proposed model society, the individual interacts through a
two-layer multiplex network, physical contact and information
spreading layers. The common nodes are affected by both the infection
and information spreading in different layers. The dynamics of
infection and information spreads are controlled by the SIR and Bass
models respectively. Adoption of information changes the attitude of
the individuals; awareness diffusion creates a group of self-protected
individuals. In this model, two types of self-protection are
considered.  Vaccination is the ultimate immunization method for most
of the viral infections. Nevertheless, vaccination requires some
effort, time and expenditure.  Hence, the first assumption is that
only $20\%$ of the population consider vaccination. The rest try to
avoid contacts with neighbors. The price of not being vaccinated is
the that the precautions, apart from the vaccination provide only
partial protection. As a second assumption, protection level of $75\%$
is used to change the characteristics of the infection spread.
Different topologies of contact and information networks embed
different diffusion dynamics. Even the same topology with increasing
number of an average degree changes the spread rates of information
and contamination. The effect of awareness on suppressing the
infection spread makes its impact if the contact network diffusion
speed is less than the spreading speed of information. The individual
responce to an epidemic stuation exhibit similarities but also vary
from the adoption of an innovation. In the epidemic case there exists
an imediate danger to the well being of the individual. Identification
of the individual responce parameters, by using the real data, may
impove the epidemic prevention efforts considerably.

\end{document}